\documentclass{osa-article}

\journal{oe}

\begin{document}

\title{Quantum noise limited nanoparticle detection with exposed-core fiber}

\author{Nicolas P. Mauranyapin,\authormark{1,*} Lars S. Madsen,\authormark{1}, Larnii Booth,\authormark{1} Lu Peng,\authormark{2,3} Stephen C. Warren-Smith,\authormark{2,3} Erik Schartner,\authormark{2,3} Heike Ebendorff-Heidepriem,\authormark{2,3}  and Warwick P. Bowen\authormark{1}}

\address{\authormark{1}ARC Centre for Engineered Quantum Systems (EQUS), School of Mathematics and Physics, The University of Queensland, Australia\\
\authormark{2}Institute for Photonics and Advanced Sensing (IPAS), School of Physical Sciences, The University of Adelaide, Adelaide, South Australia 5005, Australia\\
\authormark{3}ARC Centre for Nanoscale Biophotonics (CNBP), The University of Adelaide, Adelaide, South Australia 5005, Australia}

\email{\authormark{*}n.mauranyapin@uq.edu.au}

\begin{abstract}
Label-free biosensors are important tools for clinical diagnostics and for studying biology at the single molecule level. The development of optical label-free sensors has allowed extreme sensitivity, but can expose the biological sample to photodamage. Moreover, the fragility and complexity of these sensors can be prohibitive to applications. To overcome these problems, we develop a quantum noise limited exposed-core fiber sensor providing robust platform for label-free biosensing with a natural path toward microfluidic integration. We demonstrate the detection of single nanoparticles down to 25~nm in radius with optical intensities beneath known biophysical damage thresholds.
\end{abstract}

%%%%%%%%%%%%%%%%%%%%%%%%%%  body  %%%%%%%%%%%%%%%%%%%%%%%%%%

\section{Introduction}

% Backgroung
The past few decades have seen rapid improvements in the sensitivity of label-free optical biosensors. Single proteins of a few nanometres in size and below can now be routinely detected in solution \cite{baaske2014single,mauranyapin2017evanescent,pang2011optical,faez2015fast}. Recently, these biosensors have been used to shed light on fundamental biophysics phenomena \cite{kim2017label,baaske2014single}  and have been shown to have potential applications ranging from medical diagnostics \cite{mauranyapin2017evanescent} to high-resolution imaging \cite{baaske2016optical} and environmental monitoring \cite{baaske2014single}. Among the currently developed sensors are whispering gallery mode (WGM) resonators \cite{swaim2013detection,arnold2009whispering} and plasmonic sensors \cite{pang2011optical,chen2011enhanced}. Combining both approaches allows resolution of the motion of single enzymes \cite{kim2017label} and detection of single ions in solution \cite{baaske2014single}. 
However, these sensors generally operate with optical intensities far above known thresholds for photochemical intrusion and photodamage \cite{mirsaidov2008optimal,landry2009characterization,waldchen2015light}.
To overcome this problem, quantum-noise limited measurement techniques have been developed using a nanofiber combined with dark field heterodyne measurement \cite{mauranyapin2017evanescent}. Sensors based on these techniques are capable of detecting single BSA molecules  (3.5 nm) with four orders of magnitude reduced optical intensity. 
%Problem
However, the relative fragility of these sensors and the lack of integrated microfluidic channels has, so far, prevented their integrated used. 
%Especially for nanofibers, their small size and the difficulties in fabrication reproducibility make them hard to comply with scalability. Therefore, a more robust approach is desirable.
Nanofluidic optical fibers can overcome some of these issues and have been used to detect and differentiate label-free single viruses via their optical scattering when illuminated by the guided mode of the fiber \cite{faez2015fast,ma2016label}. 
%Proposed solution
Here we introduce a biosensor combining the advantages of these techniques using exposed-core fiber and quantum noise limited dark field heterodyne measurement. Exposed-core fibers can be interated with microfluidics channels \cite{luan2016surface} and are more robust than nanofibers due to their hundred microns diameter. We demonstrate detection of single label-free nanoparticles in solution at the quantum noise limit and compare our results with theory and simulations.

\section{Approach}

\subsection{Exposed-core fibers}

\begin{figure}[t]
 \centering
 \includegraphics[width=\linewidth]{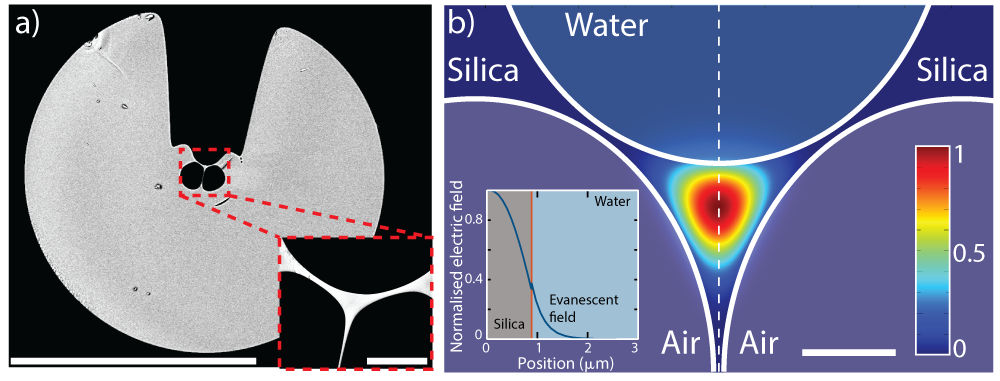}
 \caption{a) SEM image of the cross-section of an exposed-core fiber (scale bar = 100~$\mu$m). Inset: SEM picture of the core of the exposed-core fiber (scale bar = 5~$\mu$m). b) Finite element simulation of the first  fundamental propagating mode of a 2~$\mu$m diameter silica exposed-core fiber (scale bar = 2~$\mu$m). The bottom (internal) holes are filled with air and the top (open) hole is filled with water. Inset: normalized electric field as function of the distance from the center of the exposed core fiber. The inset is computed by taking a cut of b) along the dashed line.} 
 \label{mode}
 \end{figure}

 The exposed-core fibers used here are microstructured optical fibers that confine light through three longitudinal air holes \cite{schartner2017fabrication,kostecki2012silica,kostecki2014predicting}, as seen in the scanning electron microscopy (SEM) image in Fig.\ref{mode}a. The air holes introduce a large refractive index mismatch so that the light can be guided in a micrometre sized core of the same material as the cladding (see Fig.\ref{mode}b). In exposed-core fibers, one of the air holes is engineered to be open to the surrounding medium allowing direct interaction between the guided light and the environment. This is shown in Fig.\ref{mode}b where we have simulated the fundamental mode shape of a 2~$\mu$m diameter exposed-core fiber with the top hole opened to a solution of water, which could contain the particles of interest. One can see that most of the light is guided in the core of the fiber but a fraction propagates in the surrounding medium as an evanescent field (see Fig. \ref{mode}b inset). Similar to other evanescent field techniques such as WGM or nanofiber sensors, any perturbation in refractive index within the evanescent field will perturb the propagating field and can potentially be detected.
The region where the evanescent field propagates determines the interaction volume between the light and the surrounding medium. This accessible interaction volume allowed a wide variety of applications including sensors for biology \cite{li2018high,nguyen2015interferometric}. For example, exposed-core fibers have been used for the detection of fluorescence signals \cite{warren2010distributed,kostecki2014novel} and for detection of label-free molecule ensembles in an interferometric configuration \cite{li2018high,nguyen2015interferometric}. 

Compared with optical nanofibers, the advantages of exposed-core fibers are that they are made from a single material and can be consistently drawn for hundreds of meters, allowing large-scale production \cite{kostecki2012silica}. By comparaison, nanofibers must be pulled individually and their typical sensing region is only a few mm long \cite{madsen2016nondestructive}. Even though the core of the exposed-core fiber is small ($\sim$2~$\mu$m), when including the cladding they are about $120$~$\mu$m in diameter (see Fig.\ref{mode}a) making them much less fragile than nanofibers ($\sim$500~nm). Moreover, they have already been demonstrated to be compatible with microfluidics channels \cite{luan2016surface}.

%Therefore, since exposed-core fibers have been demonstrated to be a powerful tool for biosensing and can perform the same optical evanescent measurement as nanofibers (see next section) but are more robust and more scalable, they are good candidates to replace nanofibers.

\subsection{Evanescent sensing mechanism}
\label{mechanism}

\begin{figure}[t]
 \centering
 \includegraphics[width=\linewidth]{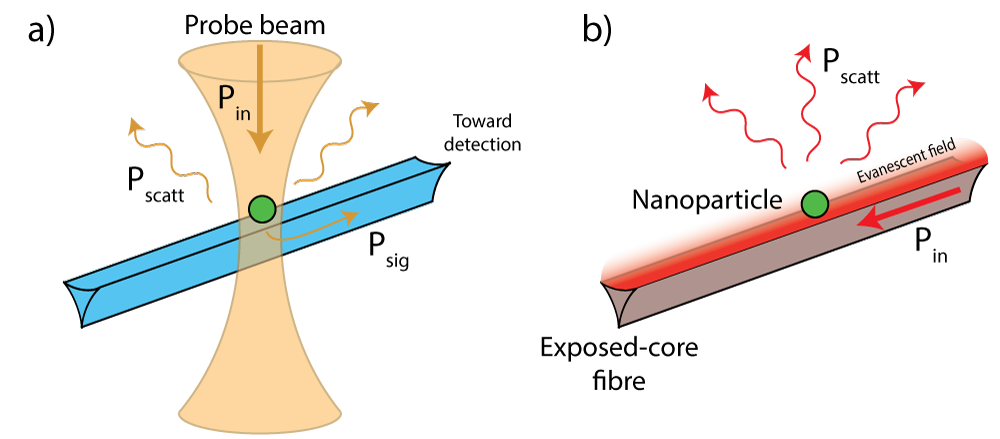}
 \caption{a) Scheme of the exposed-core sensor in dark field configuration. The incident light is sent to the nanoparticle (green sphere) on the exposed-core fiber from the top as a probe beam (yellow beam). b) Scheme of the reverse path of the light. The incident light is guided in the exposed-core fiber and the nanoparticle scatters this light out in the surrounding medium.} 
 \label{scheme}
 \end{figure}

As displayed in Fig. \ref{scheme} a), to detect nanoparticles with exposed-core fiber, we focus a probe beam (yellow) on the exposed-core fiber immersed in a solution containing nanoparticles. When a particle enters the region close to the fiber where the probe light is focussed, it will scatter the probe light and a portion of the scattered light will be collected by the guided mode of the exposed-core fiber. This forms a dark field configuration since the probe field  is orthogonal to the direction of propagation in the fiber and when no particle is present, a very small amount of probe light will be collected (additional details in  appendix section \ref{ECfiberSimulation}).

When a detection event occurs, for a focussed Gaussian probe beam, the scattered power $P_{scatt}$ by the nanoparticle is given by \cite{jacksonBook}:
\begin{equation}
	P_{scatt} = \frac{\sigma}{4 \pi w^2} P_{in},
	\label{Pscatt}
\end{equation}
where $P_{in}$ is the input power, $w$ is the probe beam waist, $\sigma$ is the scattering cross-section of the nanoparticle and we have neglected modifications of the optical density of states due to the presence of the fiber, which we expect to be small. The standard model used to determine $\sigma$ for particles smaller than the wavelength of the light is the dipole scattering model \cite{taylor2016quantum}:
\begin{equation}
	\sigma = \frac{8\pi k^4 a^6  }{3} \left(\frac{m^2-1}{m^2+2}\right)^2,
	\label{sigma}
\end{equation}
where $a$ is the radius of the particle, $k=2\pi/\lambda$ is the wave vector of the light, $\lambda$ is the wavelength of the light and $m = n_p/n_m$ is the refractive index  ratio of the nanoparticle material $n_p$ and the surrounding medium $n_m$.
The power of the detected signal at one end of the exposed-core fiber, $P_{sig}$, will depend on how much of $P_{scatt}$ is collected by the exposed-core fiber. Thus, we have $P_{sig} = \eta P_{scat}$, with $\eta$ being the collection efficiency of the exposed-core fiber.
Overall, the detection of single nanoparticles is challenging because the fraction of collected photons and the number of scattered photons can be very small. Indeed, as seen in eq. \ref{Pscatt} and \ref{sigma}  the signal power scales with the particle volume squared and is greatly reduced when the radius reaches nanometres in size.

The parameter $\eta$ is equal to the overlap integral between the scattered field and the guided modes of the exposed-core fiber, which is hard to determine analytically. However, it is closely related to the evanescent field  outside of the fiber. This can be understood using Helmholtz reciprocity. Imagine the reverse path of the light, where the incident light enters through the exposed-core fiber and is scattered out of the core by a nanoparticle on the surface of the fiber, as in Fig. \ref{scheme} b). From eq. \ref{Pscatt} we see that the power scattered will be proportional to the incident power and therefore, the stronger the electric field is at the fiber surface, the larger the power scattered. Helmoltz reciprocity tells us that $\eta$ must be the same for the forward and reverse process. Therefore, the stronger the evanescent field at the interface is, the higher the collection efficiency. 
Using finite element simulation, the collection efficiency can be calculated (see appendix in section \ref{ECfiberSimulation}). In the case of a 2~$\mu$m exposed-core fiber, we estimate the collection efficiency to be around 1 to 8 percent.

\subsection{Apparatus}

\begin{figure}[t]
 \centering
 \includegraphics[width=\linewidth]{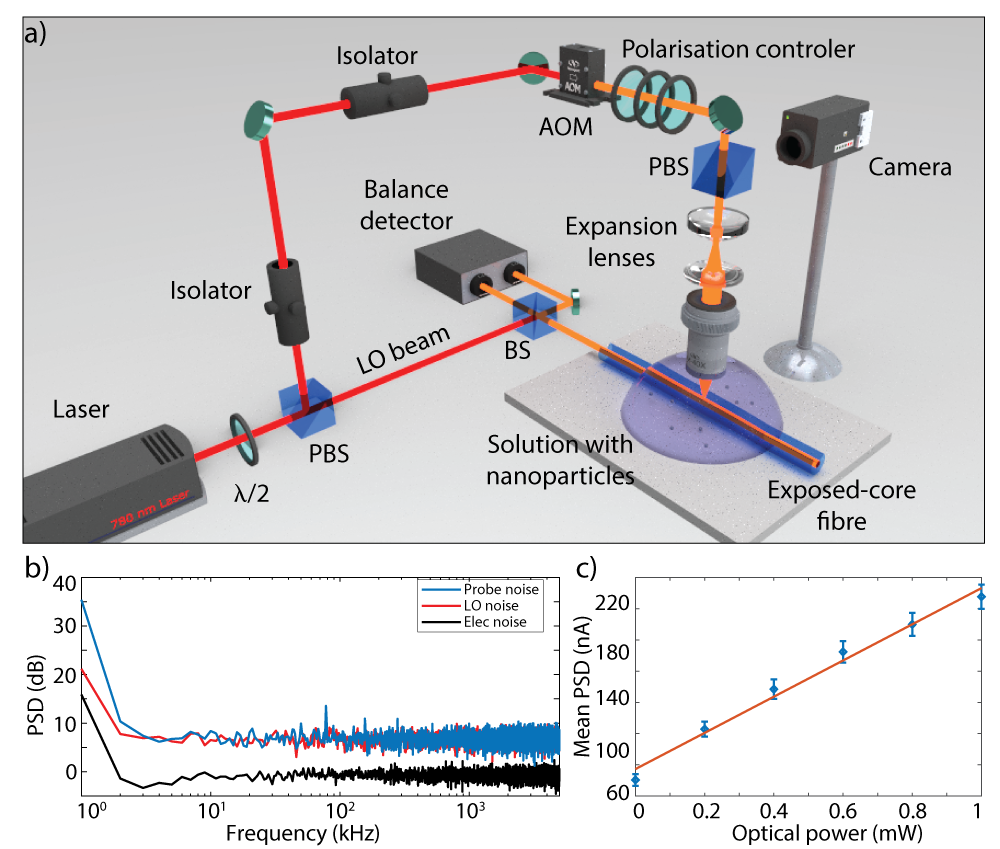}
 \caption{\label{setup}  a) Experimental setup where (P)BS stands for (polarized) beam splitter, LO for local oscillator, $\lambda/2$ for half wave plate and AOM for acousto-optic modulator.
 Two optical isolators are used to suppress any back reflected probe light to avoid contaminating the LO beam and create extra noise. The camera is used to align the objective on the fiber visualized through the polarized beam splitter situated above the microscope objective. A polarisation controller is used in the probe arm to maximize the probe power transmitted trough the PBS. Both ends of the exposed-core fiber are spiced to ultra high NA fiber to prevent the solution entering into the two air holes and the detection system.
 Note that the beam splitter before detection is in fiber but was represented in free space here for clarity.
 b) PSD of the different noises in the apparatus. The black, red and blue curves represent the electronic noise , LO noise and probe noise respectively. c) Mean power spectral density over 10~kHz of the laser noise as function of optical power sent to the detector. The blue dots represent the experimental data with error bars as the standard error of the power spectral density over 10 measurements. The red curve represents a linear fit to the experimental data. } 
 \end{figure}

Experimentally, to setup a dark field heterodyne measurement, we split a 780~nm laser beam into a local oscillator (LO) and an illumination field (probe), as seen in Fig. \ref{setup}a. The illumination field, is shifted up in frequency by an acousto-optic modulator (AOM), expanded and focused using a water immersion microscope objective (60x, 0.9 NA) on a 5 to 10~cm long exposed-core fiber. To guide the collected signal toward detection, one of the ends of the exposed-core fiber is spliced to an ultra high NA fiber (UHNA4). It is then interfered with the local oscillator on a 50-50 fiber beam splitter single mode at 780~nm. The interference is detected on a balanced detector which creates a photocurrent with a beat note at the AOM frequency. The amplitude of this beat note is proportional to the nanoparticle scattered field amplitude and is amplified by the LO. A half wave plate is used before splitting the laser to control how much power goes in each beam. Typically, 2~mW of probe power is sent to the objective and 1~mW of LO is sent to each photodiode of the balanced detector. 
Note that compared to our previous approach used in Ref \cite{mauranyapin2017evanescent}, the LO is not sent through the sensing fiber but interfered after on the beam splitter. This improves the signal to noise ratio by a factor of two since the signal is detected by both photo-diodes of the detector, and reduces the intensity of the light which the sample is exposed to.

To test the noise performance of the sensor we measured the power spectral density (PSD) of the apparatus in different configurations, see Fig. \ref{setup}b. In this figure, the black curve represents the electronic noise, with no light on the detector, the red curve the LO noise, when only the LO is detected. The blue curve represents the noise power spectral density when both probe and LO noise are sent to the detector with no particles present in the solution and the probe beam focussed on a static scatterer on the fiber. One can see that for frequencies smaller than $\sim$4~Hz the measurement is limited by electronic pick up and classical noise from the laser light. This can prevent observation of signal scattered by nanoparticles.
However, above 4~Hz, the system is not limited by the electronic noise and only the laser noise dominates. 
%A constant laser noise power for all frequencies is a signature of quantum noise and one can see that even when the probe light is focused on the exposed-core fiber, the noise power stays constant showing that the measurement is quantum noise limited for this frequency range.
One signature of the quantum laser noise is that its power spectral density is linearly proportional to the optical power due to the addition of vacuum noise \cite{mauranyapin2017evanescent}. We test this dependence in Fig. \ref{setup}c, which shows the average power spectral density of the laser noise over a bandwidth of 10~kHz against the optical power on the photodiodes of the balanced detector. The observed linear relationship shows that the laser noise is only composed of quantum noise.
This demonstrates that by using heterodyne detection, we can achieve quantum noise limited detection for most frequencies of interest, improving the signal to noise ratio and allowing better sensitivity than conventional direct detection methods.

After detection, this quantum noise limited signal is then mixed down with a home-built dual-phase lock-in amplifier (not shown on the figure) creating two orthogonal quadratures ($X$ and $Y$), which are digitalized on an oscilloscope. These two quadratures are used to extract a value $A$ proportional to the amplitude  of the collected signal in post-processing as $A=\sqrt{X^2+Y^2}$.

\section{Results}

\subsection{Nanoparticles detection}

We have tested the sensor with solutions containing either silica nanoparticles of 25~nm or 50~nm in radius or polystyrene nanoparticles of 100~nm in radius.  The solvent used here is  dulbecco's phosphate buffered saline (DPBS). DPBS was used instead of deionized water because it contains many ions that can screen the surface charges present on the exposed-core fiber and the nanoparticles. These surface charges could prevent the nanoparticles from entering the fiber groove and diffusing close to the core of the fiber due to electrostatic repulsion.

Experiments were performed following three steps. First, the exposed-core fiber was immersed in a $\sim$0.6~mL droplet of pure DPBS and the probe field was focussed on the fiber using the microscope objective. Second, a set of noise measurements were taken to record the electronic noise and the laser noise of the apparatus. Third, 40~$\mu$L of nanoparticle solution, at concentration of $3.7 \times 10^9$, $2.4 \times 10^9$ and $7.6 \times 10^9$ particles per mL for the 25~nm, 50~nm and 100~nm particles receptively, was added to the droplet and the signal was monitored and recorded on an oscilloscope.

\begin{figure}[t]
 \centering
 \includegraphics[width=\linewidth]{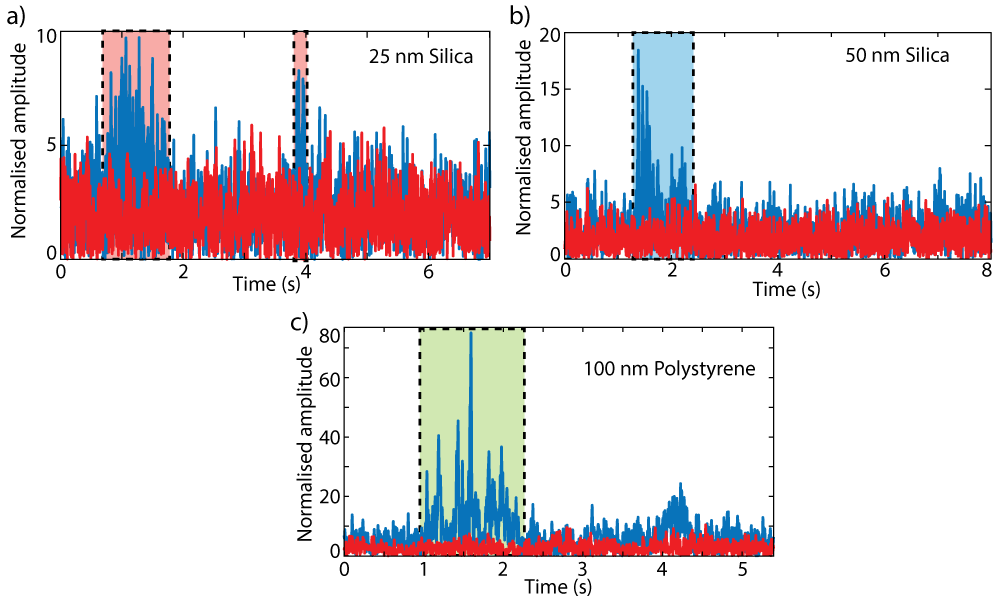}
 \caption{\label{results}  a), b), c) Time trace in blue containing detection events of single 25~nm, 50~nm silica particles and 100~nm polystyrene particles highlighted in red, blue and green respectively. The red curve in each graph represents a quantum noise trace taken before particles where added to the solution. For clarity, the traces are band pass filtered at 4~Hz to 100~Hz (quantum noise limited frequencies) and the signal amplitude is normalized by the standard deviation of the quantum noise. Before each experiment, the nanoparticle solution was tested in a Zetasizer to verify that there was no contamination or aggregations.
 } 
 \end{figure}

Typical experimental time traces are shown in Fig. \ref{results}a, b, c where the detection events are highlighted in red, blue and green for the 25~nm, 50~nm and 100~nm particles respectively. As expected (see the previous section \ref{mechanism}), larger particles have higher scattering cross-section. Consistent with this, we observe that the signal-to-noise ratio increases when increasing the nanoparticle size.

The capacity to detect 25~nm particles shows the effectiveness of heterodyne detection and compares favourably to other direct detection experiments with nanofibers which have reported detection of particles with sizes of 100~nm \cite{yu2014single} and 130~nm \cite{swaim2013tapered} in radius.
Our sensor is competitive with nanofluidic optical fibers sensors which have been demonstrated to detect single 19~nm polystyrene particles \cite{faez2015fast}. However, the accessible core, in our case, offers greater flexibility for applications, while the use of dark field heterodyne allows megahertz  detection bandwidth, compared to the camera-limited of few kilohertz in Ref.\cite{faez2015fast}.

%The 25~nm particles maximum signal is relatively low, around 8 times the standard deviation of the quantum noise on average. This shows that detection of particles smaller will be challenging with the current exposed-core fiber sensor. We believe that it is due to the size difference between exposed-core fiber and nanofiber. Indeed, the core of the exposed-core fiber is around 2~$\mu$m in diameter and the core of a nanofiber is typically around 500~nm therefore, for the same wavelength of light, the portion of the evanescent field outside the core will be much smaller for an exposed-core fiber than for a nanofiber. Consequently, the electric field at the surface of the exposed-core fiber will be smaller than the electric field at the surface of the nanofiber and, as discussed in section \ref{mechanism}, the sensitivity will be smaller. In Fig.\ref{results}d we displayed the results of a finite element simulation where we extracted the field amplitude as function of distance from the center of a nanofiber (right) and an exposed-core fiber (left) for a 780~nm light. The electric field value at the surface of the exposed-core fiber is only 34 percent of the maximum field propagating compared to 74 percent at a nanofiber surface. Therefore, the sensitivity from exposed-core fiber sensor cannot be expected to be as good as the dark field heterodyne nanofiber sensor. 

\subsection{Particle radius scaling}
\label{RadiusScaling}

\begin{figure}[t]
 \centering
 \includegraphics[width=\linewidth]{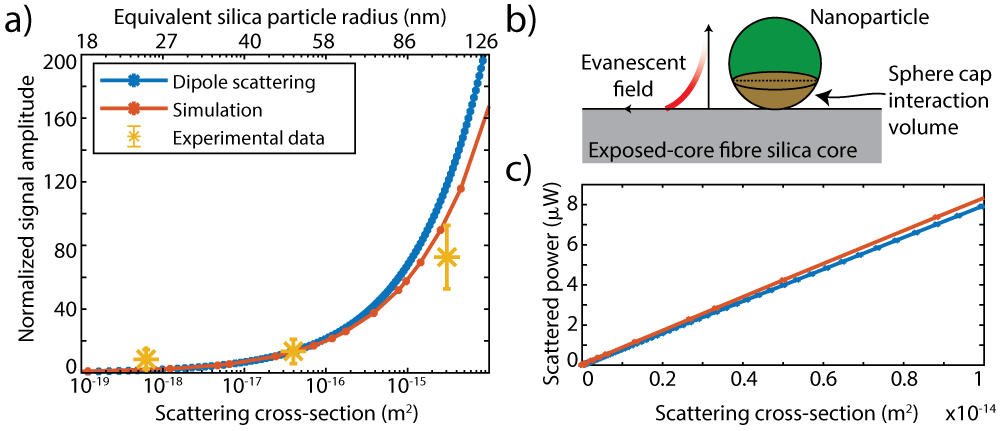}
 \caption{\label{stats}   a) Mean normalized signal amplitude against particle radius from experiment (yellow stars, error bars given by the standard deviation). This is compared to finite element simulations (red curve) and dipole scattering theory (blue curve) of the collected power. b) Schematic of the interaction between the fiber (grey) and a large nanoparticle (green).
c) Scattered power obtained from simulation (red) and theory (blue) as function of particle radius. }
 \end{figure}

In Fig. \ref{stats}a we investigate how the experimental mean maximum signal amplitude (yellow dots)  scales with particle radius. We compare the experimental data with finite element simulation of exposed-core fiber (red curve, see appendix \ref{ECfiberSimulation} for details) and with the theoretical dipole scattering of section \ref{mechanism} (blue curve). 
Because some experimental parameters such as the conversion of light signal power into voltage by the detector are unknown, the simulation and theoretical data are multiplied by a constant scaling factor to match the normalized amplitude signal of the experimental data for the 50~nm particle (see Fig. \ref{stats}a inset).
One can see that the experimental data is in relatively good agreement with the finite element simulation. However, dipole scattering overestimates the signal from particles larger than 80~nm. We expect that this is due to the exponential decay of the fiber evanescent field, such that the mean evanescent field within the volume of the nanoparticle is reduced as the particle size increases \cite{swaim2011detection}.
Indeed, for large particles, the interaction volume between the evanescent field of the fiber and the nanoparticle can be roughly modelled as a sphere dome, or sphere cap (see Fig. \ref{stats}b). The volume of this sphere cap is proportional to the radius of the particle since the depth of the evanescent field is constant. Therefore, the scattered power deviates from the dipole scattering model. 
%In other words, the collection efficiency of the fiber depends on the nanoparticle radius and we expect that the amplitude of the signal will first follow dipole scattering ($a^3 $) and then, similarly to the evanescent field , will exponentially transition to a scaling with the square root of the radius ($a^{1/2} $).

The decay length $\gamma$ of the evanescent field is equal to $\lambda/2\pi n_m $ \cite{swaim2013tapered}. In water, this decay length is equal to 93~nm for 780~nm light. From the  finite element simulation of Fig.\ref{mode}b inset, we can see that the field amplitude will decay to approximatively  4\% of the maximum field amplitude 186~nm away form the fiber surface. Hence for particles with a diameter bigger than 2$\gamma$, the signal amplitude scaling with radius will start to decrease. This is what we observe in our simulation and experiments as seen in Fig. \ref{stats}a where the dipole scattering model starts to differ for particles bigger than 80~nm. To confirm that this discrepancy does not come from errors in the simulation, we also calculated the scattered power of the nanoparticle which scales similarly to dipole scattering theory as shown in Fig. \ref{stats}b.

\section{Discussion}

We have demonstrated that exposed-core fibers combined with dark field heterodyne detection can be used to detected single label-free nanoparticles as small as 25~nm silica particles at the quantum noise limit in a biologically compatible solution. The light intensity used was only $ \sim 7 \times 10^7$~W/m$^2$ (probe beam) which is an order of magnitude smaller than intensities used for nanofibers ($ 7 \times 10^8$~W/m$^2$ in Ref.  \cite{mauranyapin2017evanescent}) and five orders of magnitude lower than intensities used in WGM resonator sensors. It is two orders of magnitude lower than known photodamage threshold \cite{mirsaidov2008optimal} showing that the exposed-core fiber sensor should be compatible with observation of biophysics phenomena for an extended period of time.
Increasing the intensity of the probe beam by an order of magnitude would still result in  intensities below photodamage thresholds and, according to eq. \ref{Pscatt},  could allow detection of single nanoparticles with a scattering cross-section ten times smaller than 25~nm silica particles with the same signal to noise ratio.
From Figs. \ref{scheme} and \ref{setup}, one can see that half of the signal is guided toward the other end of the exposed-core fiber and is not used. This signal could be also detected by using a fiber coupled mirror, for example, and could increase the signal to noise ratio of our sensor by a factor of $\sqrt{2}$.

To further enhance the sensitivity of the sensor, the exposed-core fiber could be designed with smaller core.
Our simulation in Fig. \ref{mode}b shows that the electric field amplitude at the surface of a 2~$\mu$m exposed-core fiber is around 34\% of the maximum  field amplitude. Performing a simulation with a 500~nm core exposed-core fiber shows that the field can be enhanced to 74\% of the maximum. This means that the signal amplitude could be enhanced by a factor of more than two as well as the sensitivity of the sensor assuming that the sensor remains quantum limited. In addition, the portion of the evanescent field outside the fiber will be larger, which will increase the detection volume.

In general, our results show that the exposed-core fiber sensor is sufficiently sensitive to compete with other label-free sensors. Because exposed-core fibers are less fragile and can be produced at large scale, they can be used in robust detection devices or in lab-on-chip configuration and integrated to microfluidic channels. For example, they could be used in medical diagnostic devices to detect disease markers. With such sensitivity, very few markers need to be detected and very low sample volume will be needed. 
%The robustness might also allow cleaning and re-usability of the sensor which is not possible with nanofiber.

\section{Appendix: Exposed-core fiber simulation}
\label{ECfiberSimulation}

\begin{figure}[t]
 \centering
 \includegraphics[width=\textwidth]{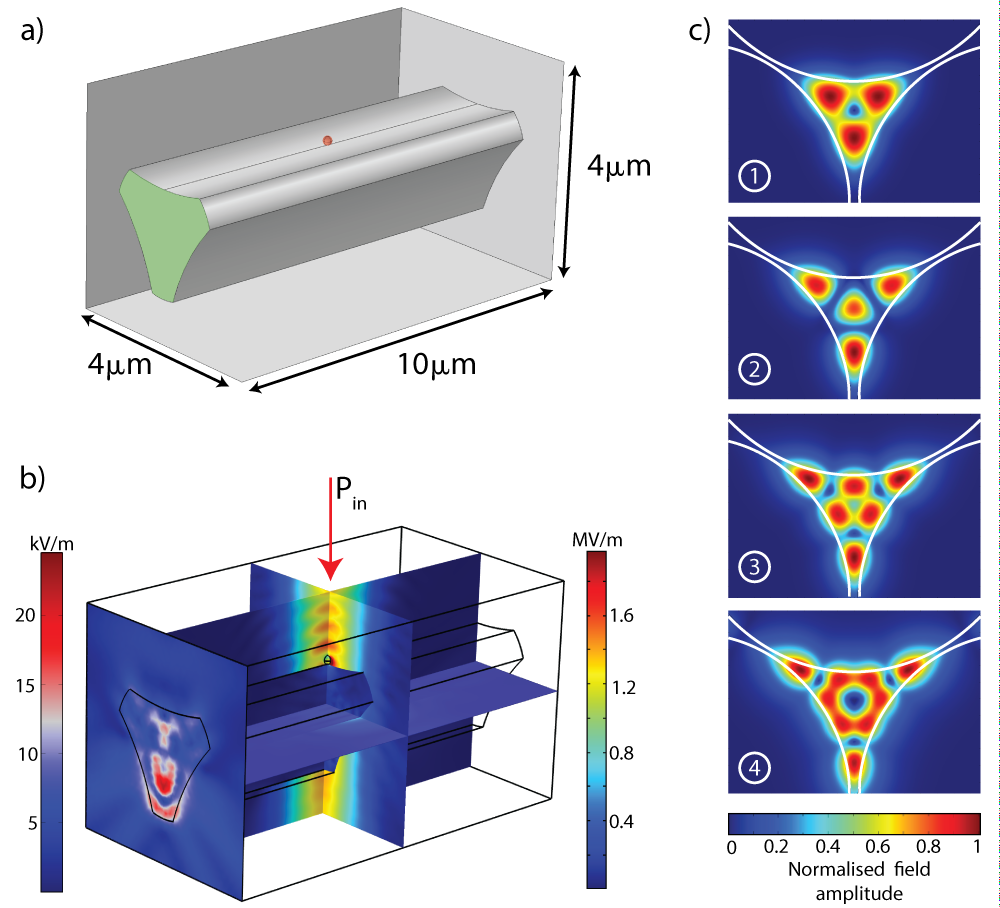}
 \caption{\label{simulation}   a) Simulation geometry showing the silica core of an exposed-core fiber in water. The green surface represents the surface used to calculated the signal power and the red sphere represents a 100~nm nanoparticle. b) Solution of the norm of the electric field. Left colorbar corresponds to the value of the field only at left end of the fiber and the right colorbar to the rest of the simulation volume. c) First four higher order modes supported by the exposed-core fiber at a wavelength of 780~nm. For this mode analysis, all holes are filled with water to be compared to b). Modes are sorted with decreasing effective refractive index: 1.431, 1.409, 1.389, 1.379 and 1.357  for the fundamental mode (0, similar to Fig. \ref{mode}b), modes (1), (2), (3) and (4) respectively.} 
 \end{figure}

To simulate the exposed-core fiber sensor, we used the finite element simulation software COMSOL multiphysics. For simplicity, only the core of the exposed-core fiber made of silica was simulated in a 4$\mu$m~$\times$~4$\mu$m~$\times$~10$\mu$m volume, as seen in Fig. \ref{simulation}a. The simulated volume was surrounded by a perfect matching layer (PML, not shown in the figure) to avoid contamination by back reflections. Because the PML material should be continuous with the simulation volume and made of a single material, we have immersed the core of the fiber in water and the air holes are also filled with water.
A 780~nm Gaussian beam of 1~mW was focused on a nanoparticle on the fiber from the top and the electromagnetic field distribution was solved for the scattered field in the entire simulation volume. Two parameters were extracted from this simulation for a range of nanoparticle radii. The first one is the collected signal power $P_{sig}$, calculated using:

\begin{equation}
	P_{sig} = \iint_S \vec{n}.\vec{\Pi}(x,y) dxdy,
	\label{Psig}
\end{equation}

 where $S$ is the surface at one end of the exposed-core fiber (see green surface in Fig. \ref{simulation}a, $\vec{\Pi}(x,y)$ is the Poynting vector at the cartesian coordinates $x,y$ and $\vec{n}$ is a unitary vector perpendicular to the surface $S$. 
In Fig. \ref{stats}a we have compared the normalized amplitude from the experiment with the amplitude of the collected signal power  $A_{sig} = C\sqrt{ P_{sig} }$ with $C$ a constant as explained in section \ref{RadiusScaling}.
The second parameter extracted from the simulation is the scattered power $P_{scatt}$ from the nanoparticle which is obtain similarly by integrating the Poynting vector around the nanoparticle surface (red surface in Fig. \ref{simulation}a).

In Fig. \ref{simulation}b, the solution of the electric field norm is shown for a 100~nm silica particle. One can see that part of the incident field is reflected back to the top by the fiber and creates some interferences but most of the field is transmitted and exits the simulation volume from the bottom. The solution of the collected signal field is displayed by the cut at the left end of the fiber. One can see that the collected light is not only guided in the fundamental mode of the exposed-core fiber displayed in Fig. \ref{mode}b. We believe the nanoparticle excites several modes supported by the exposed core fiber and the the cut of collected signal results from the interference between these modes. The first few modes are displayed in Fig. \ref{simulation}c and one can see that the collected signal could be mostly supported by the second and fourth mode.

Using eq. \ref{Psig}, the collected signal power calculated for a 100~nm silica particle is 5.3$\times$10$^{-4}$~mW and the scattered power is around 7.4$\times$10$^{-3}$~mW. From these two parameters, the collection efficiency can be calculated using $\eta = P_{sig}/P_{scatt}$ which gives 7.2 percent for the 100~nm particle. For particles of other sizes, ranging from 10~nm to 200~nm, the calculated collection efficiency is between 1.7 and 8.1 percent.
Note that the collection efficiency is actually twice this value as there are collected photons that propagate in the other direction, but are not detected. 

Finally, to verify that our simulation is valid, a control simulation was performed with no particle and showed that the collected power at the end of the fiber was at least 6 orders of magnitude lower than when a 100~nm particle is present.

\paragraph{Funding:} This work was supported by a grant from the Air Force Office of Scientific Research (grant no. FA9550-17-10397).  The authors would like to acknowledge ARC Centre of Excellence for Engineered Quantum Systems (CE110001013), ARC Centre of Excellence for Nanoscale Biophotonics (CE14010003) and the OptoFab node of the Australian National Fabrication Facility utilizing Commonwealth and South Australian State Government funding. W.P.B. acknowledges a fellowship from the Australian Research Council (FT140100650). S.C.W-S. acknowledges a Ramsay Fellowship from the University of Adelaide. L.P. acknowledges financial support from the China Scholarship Council.

\paragraph{Acknowledgments:} The authors thank Alastair Dowler and Evan Johnson from the University of Adelaide for the exposed-core fiber fabrication.

\end{document}